\documentclass[aps,twocolumn,superscriptaddress,altaffilletter,lengthcheck,
tightenlines,showpacs,showkeys]{revtex4}
\usepackage{multirow}
\newcommand{\al}{\alpha}
\newcommand{\bb}{\beta}
\newcommand{\D}{\Delta}
\newcommand{\ben}{\begin{eqnarray}}
\newcommand{\een}{\end{eqnarray}}
\newcommand{\be}{\begin{equation}}
\newcommand{\ee}{\end{equation}}
\newcommand{\ba}{\begin{eqnarray}}
\newcommand{\ea}{\end{eqnarray}}
\newcommand{\n}{\label}

\newcommand{\ro}{\rho}

\newcommand{\bn}{\begin{equation}\label}

\usepackage[dvipdf]{epsfig}
\usepackage{color}
\begin{document}
\title{Interacting dark matter and modified holographic Ricci dark energy\\
plus a noninteracting cosmic component\\  }
\author{Luis P. Chimento}\email{chimento@df.uba.ar}
\affiliation{Departamento de F\'{\i}sica, Facultad de Ciencias Exactas y Naturales,  Universidad de Buenos Aires, Ciudad Universitaria, Pabell\'on I, 1428 Buenos Aires, Argentina}
\author{Mart\'{\i}n G. Richarte}\email{martin@df.uba.ar}
\affiliation{Departamento de F\'{\i}sica, Facultad de Ciencias Exactas y Naturales,  Universidad de Buenos Aires, Ciudad Universitaria, Pabell\'on I, 1428 Buenos Aires, Argentina}
\date{\today}
\bibliographystyle{plain}
\begin{abstract}
We investigate a spatially flat Friedmann-Robertson-Walker  universe that has an interacting dark matter, a modified holographic Ricci dark energy (MHRDE), plus a third, decoupled component that behaves as a radiation term. We consider a nonlinear interaction in the dark component densities and their derivatives up to second order. We apply the $\chi^{2}$ method to the observational Hubble data for constraining the cosmological parameters and  analyze the amount of dark energy in the radiation era  for both MHRDE and holographic Ricci dark energy  models. The former is  consistent with the bound $\Omega_{x}(z\simeq 1100)<0.1$ reported for the behavior of dark energy at early times while the latter  does not fulfill it.
\end{abstract} 
\vskip 1cm
\bibliographystyle{plain}
\maketitle
\section{Introduction}
The current lore of modern cosmology indicates that  95 \% of the matter-energy in the 
Universe is harbored in the dark sector composed of dark matter (DM) and dark energy (DE) \cite{obse1},\cite{obse2}. Observations suggest that DM is a substantial nonbaryonic component that would be responsible for structure formation  in the Universe \cite{dme}. Likewise,  DE has been the cornerstone of modern cosmology ever since the measurements of supernova type Ia  have confirmed its central role in explaining the current speeding up of the Universe \cite{obse1}. The existence of  DM and DE have been supported by observations  such as  cosmic microwave background and power spectrum  of clustered matter among other probes \cite{obse1,obse2, dme, obse3}.  Here we will work within the framework of dynamical DE by taking into account as DE ansatz, $\ro_{x}$, the modified holographic Ricci dark energy (MHRDE)  based on the holographic principle \cite{holop}. The holographic DE model assumes that the quantum zero point energy density $\ro_{\Lambda}=3c^{2}M^{2}_{P}/L^{2}$ is equal to DE density $\ro_{x}$ \cite{hc}, being $L$, an infrared cutoff that will be related with a cosmological length \cite{hl1}, \cite{hl2}, \cite{hl3a}, \cite{hl3b}, \cite{holoG}. 

Despite the lack of microscopic theory to describe  DM and DE, their existence offers some of the most compelling evidence for physics beyond the standard paradigm.  However there is an unavoidable degeneracy between DM and DE within Einstein's gravity. There could be a hidden nongravitational coupling between them without violating current observational constraints, and thus, it is interesting to develop ways of testing different kinds of interactions in the dark sector. 

The fraction of DE at recombination era should fulfill the bound $\Omega_{x}(z\simeq 1100)<0.1$ in order for the DE model be consistent with the big bang nucleosynthesis (BBN) data.  Uncovering the nature of DE. as well as its properties to high redshift, could serve as an invaluable guide to the physics behind the recent speed up of the Universe \cite{EDE1}. Current and future data were examined for constraining the amount  of dark energy at early times, and the data analyzed has led to an upper bound of  $\Omega_{e}<0.043$ with $95\%$ confidence level (C.L.) in case of relativistic early dark energy, while for a quintessence type of EDE has given $\Omega_{e}<0.024$, and  although   the EDE component is not preferred, it is also not 
excluded from the current data \cite{EDE1}. Another forecast for  the amount of EDE  was obtained with the  Planck  and CMBPol experiments \cite{EDE2}; by studying  the stability of the value $\Omega_{x}(a \simeq 10^{-3}) \simeq 0.03$, it was found that $1\sigma$ error coming from Planck experiment is $\sigma^{Planck}_{e} \simeq 0.004$, whereas the CMBPol improved this bound by a factor of 4 \cite{EDE2}. 

In the present work we investigate a Universe  that has an interacting dark sector, and  a decoupled component that could mimic a radiation term.  We constrain the cosmic set of parameters by using the Hubble data and  the severe bounds found for  DE at early times.
\section{The model}
We will apply the holographic principle \cite{holop} to a cosmological model \cite{hc} associating the infrared cutoff $L$ with the modified Ricci radius, thus we take $L^{-2}$ in the form of a linear combination of $\dot H$ and $H^2$. After that, the MHRDE $\rho_{x}=3c^{2}M^{2}_{~P}L^{-2}$ \cite{hl1}, \cite{hl2}, \cite{hl3a}, \cite{hl3b}, becomes 
\be
\n{03}
\ro_x=\frac{2}{\al -\bb}\left(\dot H + \frac{3\al}{2} H^2\right).
\ee
Here $H = \dot a/a$ is the Hubble expansion rate, $a$ is the scale factor and $\al, \bb$ are free constants. Introducing the variable $\eta = \ln(a/a_0)^{3}$, with $a_0$ the present value of the scale factor and $' \equiv d/d\eta$, the above MHRDE (\ref{03}) looks like a modified conservation equation (MCE) for both cold DM $\ro_c$ and the MHRDE with constant equations of state plus a  continuity equation for the decoupled  component named  $\ro_{m}$,
\be
\n{06}
\ro'=-\al\ro_c -\bb\ro_x,\quad  \ro_{m}'+ (\omega_m +1)\ro_{m}=0,
\ee
after using the Friedmann equation, 
\be
\n{00}
3H^2= \ro +\ro_{m},
\ee 
where $\ro=\ro_c + \ro_x$ is total energy density of the interacting dark sector and $\ro_{m}=\ro_{m0}a^{-3(\omega_m +1)}$. In connection with observations on the large scale structures, which seems to indicate that the Universe  must have been dominated by nearly pressureless components, we assume that $\ro_c$ includes all these components and has an equation of state (EOS) $p_c=0$ while for $\ro_{x}$ and $\ro_{m}$ we have  $\omega_x=p_x/ \ro_x$ and $\omega_m=p_m/ \ro_m$, respectively. So that the whole conservation equation (WCE) for the dark sector becomes $\ro'=-\ro_c -(\omega_x +1)\ro_x$. In addition, the compatibility between the MCE (\ref{06}) and the WCE yields a linear dependence of the equation of state of the MHRDE with  $r=\ro_c/\ro_x$, $\omega_x=(\al -1)r+ (\beta-1)$. Solving the linear algebraic system of equations (\ref{06}) and $\ro=\ro_c+\ro_x$, we acquire both DE densities as functions of $\ro$ and $\ro'$,
\be
\n{10}
\ro_c= - \frac{\bb \ro +\ro '}{\D}, \qquad \ro_x= \frac{\al \ro +\ro '}{\D},
\ee
with $\D = \al -\bb$, while the total pressure of the dark mix is $p = p_c + p_x= -\ro -\ro'$. From now on we will use the  simpler MCE (\ref{06}) instead of the WCE with variable $\omega_{x}$,  and introduce an interaction between both dark components through the term $3HQ$ into the MCE (\ref{06}), 
\be
\n{08}
\ro_c' + \al \ro_c = - Q,\quad \ro_x' + \bb \ro_x = Q.
\ee
From Eqs. (\ref{10}) and (\ref{08}), we obtain the source equation \cite{jefe1} for the energy density $\ro$,
\be
\n{14}
\ro''+(\al + \bb)\ro' + \al\bb\ro =  Q\D,
\ee
where the interaction $Q$ between both dark components is nonlinear, and includes a set of terms  which are the total energy density of the dark sector,  its first and second derivatives\cite{jefe1}, 
\be
\n{qnl}
Q=\frac{(1-n)\Gamma}{\Delta\gamma}\frac{\ro'^{2}}{\ro}+\frac{\al\beta}{\Delta}\ro + \frac{[\al+\beta-n\Gamma (1+\delta)]}{\Delta}\ro'+\frac{(1-\Gamma)}{\Delta}\ro''
\ee
being $n$, $\delta$, and $\Gamma$ positive constants.  It turns out to be that the interaction (\ref{qnl}) gives rise to an effective fluid whose EOS includes all known Chaplygin gases \cite{jefe1}. In particular,  it generates a new broad kind of gas that was called relaxed Chaplygin gas, \cite{jefe1}, \cite{hl3b}. The aim of this work is  to explore a model, which is  characterized by an effective fluid  having  an EOS of a new kind of Chaplygin gas encoded in Eq. (\ref{qnl}) by using observational constraints coming from the behavior of dark energy at early times.

Replacing Eq.(\ref{qnl}) into Eq.(\ref{14}) turns into a nonlinear second order differential equation for the total energy density $\ro$,  a.k.a source equation (see \cite{jefe1} for further details). Introducing the new variable $y=\ro^{n}$ into the latter equation one gets a second order linear differential equation $y''+(1+\delta)n y'=0$, whose solution is $\ro=\left[Aa^{-3n(1+\delta)}+ B\right]^{1/n}$, where $A$ and $B$ are positive constants. Note that Eq.(\ref{qnl}) generalizes the EOS associated with the modified  Chaplygin gas in the sense that the exponents $1/n$ and $-3n(1+\delta)$ are  different. Plugging $\ro(a)$ in Eq. (\ref{10}), we have both dark component densities
\be
\n{cI}
\ro_c=\frac{-\ro}{\D}\left[\bb-(1+\delta)+\frac{B(1+\delta)}{\ro^{n}}\right],\,\,\,
\ee
\be
\n{xI}
\ro_x=\frac{\ro}{\D}\left[\al-(1+\delta)+\frac{B(1+\delta)}{\ro^{n}}\right],\,\,\,\,\,\,\,\,\,\,\,\,	
\ee
while the total energy density is given by 
\be
\n{ET}
\ro_{t}=\ro_{m0}a^{-3(\omega_m +1)}+ \left[Aa^{-3n(1+\delta)}+B\right]^{1/n}
\ee
From Eq.(\ref{ET}) we see that the Universe is dominated by radiation  for $\omega_{m}=1/3$ at early times where the dark components are negligible, after this epoch it enters  an era dominated by dark matter and finally  exhibits a de Sitter phase at late times.  Then, the interaction allows a smooth interpolation between a cold  matter era in the distant past (intermediate regime) to a speeding up stage
at late times. Finally,  the positivity of dark densities require to take $\beta<0$, $\al>0$ along with $\al \geq 1+\delta$. From now on we adopt the latter restrictions.

\section{Observational Hubble data  constraints}
\begin{figure}
\begin{center}
\includegraphics[height=8.5cm,width=6.6cm]{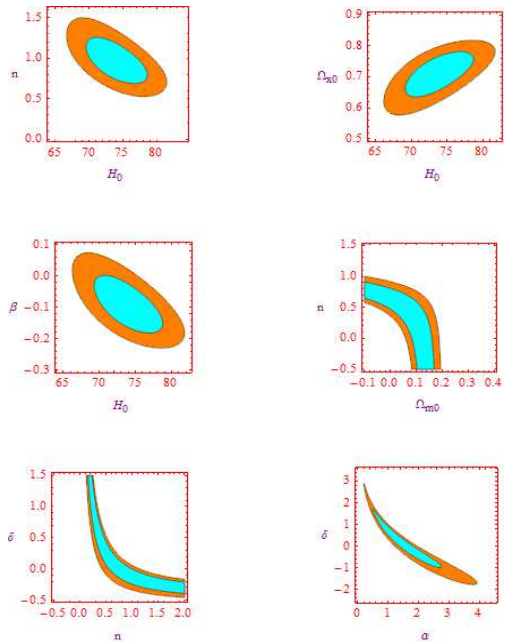}
\caption{Two-dimensional C.L. associated with $1\sigma$,  $2\sigma$ for different $\theta$ planes}.
\label{F1}
\end{center}
\end{figure}

Using the observational  $H(z)$ data \cite{obs3}, we focus on the observational constraints on the parameter space of the MHRDE  plus the third component. The  statistical analysis is based on the $\chi^{2}$ function of the Hubble data which is constructed as (see e.g.\cite{Press})
\be
\n{c1}
\chi^2(\theta) =\sum_{k=1}^{12}\frac{[H(\theta,z_k) - H_{obs}(z_k)]^2}{\sigma(z_k)^2},
\ee
where the $\theta$ parameters are the cosmological set of parameters, $H_{obs}(z_k)$ is the observational $H(z)$ data at the redshift $z_k$, and $\sigma(z_k)$ is the corresponding $1\sigma$ uncertainty \cite{obs3}.  Using Eqs. (\ref{06}), (\ref{cI}), and (\ref{xI}) one gets the Hubble parameter in term of the redshift $x=1+z$ and the relevant cosmological parameters
\be
\n{Ht2}
\frac{H(z)}{H_0}=\Big[(1-\Omega_{m0})\big(J+ (1-J)x^{3n(\delta+1)}\big)^{1/n}+\Omega_{m0}x^{3\alpha}\Big]^{1/2}
\ee
\be
\n{B2}
J[\theta]=\frac{\Omega_{x0}\D-\al(1-\Omega_{m0})}{(1+\delta)(1-\Omega_{m0})}+1
\ee
where  we have used that $B(1-J)=A J$ and the flatness condition:$1=\Omega_{x0}+\Omega_{c0}+\Omega_{m0}$. Now, we perform the statistical analysis by minimizing the $\chi^2$ function for obtaining the best-fit values of  the random variables $\theta=\{\al, \beta,  \delta, n,  \Omega_{x0}, \Omega_{m0}, H_{0}\}$.  The  best-fit parameters $\theta_{c}$ are those values where $\chi^2_{min}(\theta_{c})$ leads to the local minimum of the $\chi^2(\theta)$ distribution. If the goodness condition (GC) $\chi^2_{d.o.f}=\chi^2_{min}(\theta_{c})/(N -n) \leq 1$ is satisfied then the data are consistent with the considered model $H(z|\theta)$ \cite{Press}. We will estimate two parameters and the rest of them will be taken as priors, this will be done until all parameters have been varied [Fig.(\ref{F1})], but for simplicity we will show some of these constraints only. As usual, in the case of two parameters, the  $68.3\%$, $95.4\%$  C.L.  are made of the random data sets that satisfy the inequality $\Delta\chi^{2}=\chi^2(\theta)-\chi^{2}_{min}(\theta_{c})\leq 2.30$, $\Delta\chi^{2}\leq 6.17$ [Fig.(\ref{F1})].
\begin{figure}
\begin{center}
\includegraphics[height=7cm,width=8.5cm]{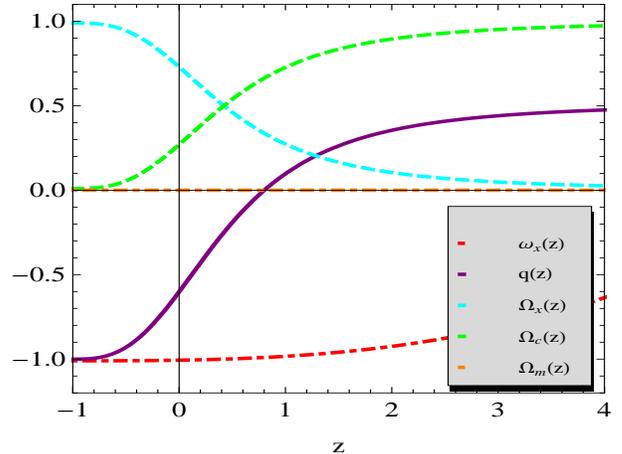}
\caption{Plot of $\omega_{x}$, $q$, $\Omega_x$, $\Omega_c$, $\Omega_m$ in terms of  the redshift $z$.}
\label{F3}
\end{center}
\end{figure}

The 2-dimensional CL obtained with standard $\chi^{2}$ function for two parameters is shown in Fig.(\ref{F1}) whereas the estimation of these parameters is briefly summarized in Table (\ref{I}). In order to contrast the MHRDE ansatz with HRDE model,  we performed the same  estimation in both cases. We find  the best-fit at  $(n, H_{0})=(0.64, 73.60 {\rm km~s^{-1}\,Mpc^{-1}})$ with $\chi^2_{d.o.f}=0.758$ by  using  the same priors of the MHRDE ansatz except for the value taken by $\al$. We also get  the best-fit at  $(\Omega_{x0}, H_{0})=(0.76,  73.81 {\rm km~s^{-1}\,Mpc^{-1}})$ along with $\chi^2_{d.o.f}=0.760$ [cf.  Table (\ref{I})]. These findings show that both models are in agreement with the standard values of  $\Omega_{x0}$ and  $H_{0}$ reported by  the WMAP-7 project \cite{WMAP7}.  Using the  pair $(n, H_{0})=(0.95, 74.09 {\rm km~s^{-1}\,Mpc^{-1}})$ [Table (\ref{I})],  we show in Fig. (\ref{F3}) the behavior of the decelerating parameter with the redshift; in particular, its present-day value is $q_{0}=-0.60$, whereas for the HRDE model gives $q_{0}=-0.59$, so both values are in agreement with the ones reported  by the WMAP-7 \cite{WMAP7}. In Fig.(\ref{F3}) we plot the density parameters $\Omega_c$, $\Omega_x$, $\Omega_c$, and find their present-day values for MHRDE model, namely, $\Omega_{x0}=0.73$, $\Omega_{m0}=10^{-8}$, $\Omega_{c0}\simeq 0.269$, whereas HRDE ansatz leads to $\Omega_{x0}=0.79$, $\Omega_{m0}=10^{-8}$, and $\Omega_{c0}\simeq 0.208$. The EOS for dark energy does not cross the phantom barrier $\omega_{x}=-1$, so the model does not exhibit a quintom phase \cite{quinto}.

We  devote our attention to the behavior of DE at early times  within the framework of interacting dark sector plus a decoupled third component. We begin by setting $\al=4/3$ to get a radiation contribution in the total density, and,  as is well known in the radiation era the stringent bound $\Omega_{x}(z\simeq 1100)<0.1$ should be fulfilled by the model in order to be consistent with data coming from the recombination era. In the case of the HRDE  model with $(\al, \beta,  \delta, n, \Omega_{x0}, \Omega_{m0}, H_{0})=(4/3,  -0.01, 0.01, 0.8,0.76, 10^{-8}, 73.81 ~{\rm km~s^{-1}\,Mpc^{-1}})$ the fraction of DE is $\Omega_{x}(z\simeq 1100)=0.24$; therefore, it does not satisfy this bound. For the MHRDE case, we take  $(\al, \beta,  \delta, n, \Omega_{x0}, \Omega_{m0}, H_{0})=(1.01,  -0.01, 0.01, 0.9,0.71, 10^{-8}, 73.94 ~{\rm km~s^{-1}\,Mpc^{-1}})$ along with $\chi^{2}_{d.o.f}=0.763< 1$. The latter case leads to an EDE with $\Omega_{x}(z\simeq 1100)=1.2 \times 10^{-8}<0.1$ that is consistent with both the bounds reported in \cite{EDE1}, and with the future constraints achievable by Planck and CMBPol experiments \cite{EDE2} because at  large redshifts we get some bounds for the DE such as $\Omega_{x}(z\simeq 10^{10})=1.3 \times 10^{-27}$, $\Omega_{x}(z\simeq 10^{16})=5.8 \times 10^{-44}$, and $\Omega_{x}(z\simeq 10^{20})=7.23 \times 10^{-55}$. Then, taking into account the third component as a radiation term or a nearly radiation contribution helped to validate  our model  concerning the present-day constraints, but also it  shows that the value of the cosmological parameters selected are consistent with BBN constraints. 

\section{conclusion}
We have considered a model of three components composed of an interacting dark sector with MHRDE and a noninteracting component. We have introduced a  nonlinear interaction such that the dark mix interpolates between a cold DM regime at early times (after a radiation era) and a de Sitter phase in the far future. 

In the case of 2D C.L., we have found that  the MHRDE model leads to the best-fit at $(\Omega_{x0}, H_{0})=(0.71, 73.94 {\rm km~s^{-1}\,Mpc^{-1}})$ with $\chi^2_{d.o.f}=0.763$ [Fig.(\ref{F1}) and Table (\ref{I})] while   the HRDE  leads to a best-fit at  $(\Omega_{x0}, H_{0})=(0.76,  73.81 {\rm km~s^{-1}\,Mpc^{-1}})$ plus a $\chi^2_{d.o.f}=0.760<1$.  Both models are in agreement with the standard  values of  $\Omega_{x0}$ and  $H_{0}$ reported by  the WMAP-7 project \cite{WMAP7} and  fulfill the GC. Using the best fit-value $(n, H_{0})=(0.95, 74.09 {\rm km~s^{-1}\,Mpc^{-1}})$ [ Table (\ref{I})], we have shown that the behavior of the $q(z)$ exhibits  an accelerated phase at late times [see Fig. (\ref{F3})] and obtained that its present-day value is $q_{0}=-0.60$, whereas for the  HRDE model yielded $q_{0}=-0.59$, so both values are in agreement with the ones reported  by the WMAP-7 \cite{WMAP7}.  Since the HRDE model with $(\al, \beta,  \delta, n, \Omega_{x0}, \Omega_{m0}, H_{0})=(4/3,  -0.01, 0.01, 0.8,0.76, 10^{-8}, 73.81 ~{\rm km~s^{-1}\,Mpc^{-1}})$ along with $\chi^{2}_{d.o.f}=0.760< 1$, leads to an EDE $\Omega_{x}(z\simeq 1100)=0.24$,  then this model  does not fulfill the bound for EDE at the recombination era. On the other hand,  for the MHRDE case we have taken $(\al, \beta,  \delta, n, \Omega_{x0}, \Omega_{m0}, H_{0})=(1.01,  -0.01, 0.01, 0.9,0.71, 10^{-8}, 73.94 ~{\rm km~s^{-1}\,Mpc^{-1}})$ along with $\chi^{2}_{d.o.f}=0.763< 1$. The latter case leads to an EDE  $\Omega_{x}(z\simeq 1100)=1.2 \times 10^{-8}<0.1$ which is consistent with  both the bounds reported in \cite{EDE1}, and with the future constraints achievable by Planck and CMBPol experiments \cite{EDE2}.
\acknowledgments
L.P.C thanks  UBA under Project No. X044 and the CONICET under Project PIP 114-200801-00328. M.G.R is supported by CONICET.

\begin{widetext}
\begin{center}
\begin{table}
\begin{tabular}{|l|l|l|}
\hline
\multicolumn{3}{|c|}{2D Confidence Level} \\
\hline
Priors & Best fits & $\chi^{2}_{dof}$\\
\hline
\multirow{1}{*}{$(\alpha,\beta,\delta,\Omega_{x0},\Omega_{m0})=(1.01, -10^{-2},10^{-3},0.73, 10^{-8})$} & $(n, H_{0})=(0.95, 74.09)$& $0.766$\\
\hline
\multirow{1}{*}{$(\alpha,\beta,\delta,n,\Omega_{m0})=(1.01, -10^{-2},10^{-3},0.9, 10^{-8})$} & $(\Omega_{x0}, H_{0})=(0.71, 73.94)$& $0.763$\\
\hline
\multirow{1}{*}{$(\alpha,\beta,\delta,\Omega_{x0}, H_{0})=(1.2, -10^{-2},0.1,0.73, 74.20)$} & $(n,\Omega_{m0})=(0.63,10^{-8})$& $0.764$\\
\hline
\multirow{1}{*}{$(\alpha,\delta,n,\Omega_{x0},\Omega_{m0})=(1.26, 0.1,0.8,0.73, 10^{-8})$} & $(\beta, H_{0})=(-0.09, 74.09)$& $0.771$\\
\hline
\multirow{1}{*}{$(\alpha,\beta,\Omega_{x0},\Omega_{m0}, H_{0})=(1.2, -10^{-2},10^{-3},0.73, 10^{-8}, 74.2)$} & $(\delta,n)=(0.1,0.63)$& $0.764$\\
\hline
\multirow{1}{*}{$(\beta,n,\Omega_{x0},\Omega_{m0}, H_{0})=(-10^{-2},0.4,0.73, 10^{-8}, 74.2)$} & $(\delta, \alpha)=(0.25, 1.37)$& $0.766$\\
\hline
\end{tabular}
\caption{\label{I}Observational bounds for the 2D C.L. obtained in Fig. (\ref{F1}) by varying two-parameters.}
\end{table}
\end{center}
\end{widetext}
\end{document}